\documentclass[11pt,fleqn,twoside]{article}
\usepackage{amsfonts,amssymb,latexsym}
\makeatletter
\newcommand{\prava}{\footnotesize\it
\begin{flushright}
\begin{minipage}{18cm}
Copyright \copyright 1998 by S.Yu. Sakovich
\end{minipage}
\end{flushright}}

\newcommand{\name}[1]{\begin{flushleft}
                       \LARGE \bf #1
                       \end{flushleft}\vspace{-3mm}}

\newcommand{\Author}[1]{\begin{flushleft}
                       \it #1 \end{flushleft}}

\newcommand{\Adress}[1]{\begin{flushleft}
                       \it #1 \end{flushleft}}

\newcommand{\Date}[1]{\begin{flushleft}
                      \small  \it #1 \end{flushleft}}

\newcommand{\ehkol}{Author \ name}
\newcommand{\ohkol}{Article \ name}
\renewcommand{\@evenhead}{
\hspace*{-3pt}\raisebox{-15pt}[\headheight][0pt]{\vbox{\hbox to \textwidth
{\thepage \hfil \ehkol}\vskip4pt \hrule}}}
\renewcommand{\@oddhead}{
\hspace*{-3pt}\raisebox{-15pt}[\headheight][0pt]{\vbox{\hbox to \textwidth
{\ohkol \hfil \thepage}\vskip4pt\hrule}}}
\renewcommand{\@evenfoot}{}
\renewcommand{\@oddfoot}{}

\newcommand{\be}{\begin{equation}}
\newcommand{\ee}{\end{equation}}
\newcommand{\ba}{\hspace*{-5pt}\begin{array}}
\newcommand{\ea}{\end{array}}

\makeatother

\begin{document}
\setcounter{page}{230}
\thispagestyle{empty}

\renewcommand{\ehkol}{S.Yu. Sakovich}
\renewcommand{\ohkol}{Integrability of a 
Perturbed KdV Equation}

\begin{flushleft}
\footnotesize \sf
Journal of Nonlinear Mathematical Physics \qquad 1998, V.5, N~3,
\pageref{sakovich-fp}--\pageref{sakovich-lp}.
\hfill {\sc Letter}
\end{flushleft}

\vspace{-5mm}

\renewcommand{\footnoterule}{}
{\renewcommand{\thefootnote}{}
 \footnote{\prava}}

\name{On Integrability of a (2+1)-Dimensional
Perturbed KdV Equation}\label{sakovich-fp}
\Author{S.Yu. SAKOVICH}

\Adress{Institute of Physics, National Academy of Sciences,
P.O. 72, Minsk, Belarus\\
E-mail: sakovich@dragon.bas-net.by}

\Date{Received April 14, 1998; Accepted June 5, 1998}

\begin{abstract}
\noindent
A (2+1)-dimensional perturbed KdV equation, recently introduced by W.X.~Ma
and B.~Fuchssteiner, is proven to pass the Painlev\'e test for integrability
well, and its 4$\times $4 Lax pair with two spectral parameters is found.
The results show that the Painlev\'e classif\/ication of coupled KdV equations
by A.~Karasu should be revised.
\end{abstract}

\bigskip

\noindent
Recently, Ma and Fuchssteiner [1] applied a new scheme of
perturbation to the KdV equation and derived the following (2+1)-dimensional
system:
\begin{equation}
u_t=u_{xxx}+6uu_x,  \label{sakovich:1}
\end{equation}
\begin{equation}
v_t=v_{xxx}+6\left( uv\right) _x+3\left( u_{xx}+u^2\right) _y.  \label{sakovich:2}
\end{equation}
They also raised the question of its integrability.

In the present letter, we verify the integrability of the system (1)--(2) by
means of the Painlev\'e test, show that this system possesses a 4$\times $4
Lax pair with two spectral parameters, and then discuss our results in
relation with one recent Painlev\'e classif\/ication of coupled KdV
equations [9].

The singularity analysis may be applied in studying the 
integrability of nonlinear dif\/ferential equations [2]. Let us apply the
Weiss-Kruskal algorithm of singularity analysis [2]--[4] to the
system (1)--(2). This system is a normal system [5] with
non-characteristic hypersurfaces $\varphi \left( x,y,t\right) =0$
satisfying the condition $\varphi _x\neq 0$
(we take $\varphi _x=1$), and its general solution should contain six
arbitrary functions of two variables each. Substituting $u=u_0\left(
y,t\right) \varphi ^\sigma +\cdots+u_r\left( y,t\right) \varphi ^{\sigma
+r}+\cdots $ and
$v=v_0\left( y,t\right) \varphi ^\tau +\cdots+v_r\left( y,t\right)
\varphi ^{\tau +r}+\cdots$ into (1)--(2), we f\/ind the admissible exponents $%
\sigma $ and $\tau $ of the dominant behavior of solutions and the
corresponding positions $r$ of resonances. There is only one branch to be
analyzed: $\sigma =-2,$ $\tau =-3,$ $u_0=-2,$ $\forall \; v_0\left(
y,t\right) , $ $r=\underline{-1},0,\underline{4},5,\underline{6},7$
(underlined are the
positions of resonances of the KdV equation (1); all other branches are
either special cases of this branch or Taylor expansions governed by the
Cauchy-Kovalevskaya theorem). Then we substitute
$u=\sum\limits_{i=0}^\infty u_i(y,t)\varphi ^{i-2}$ and
$v=\sum\limits_{i=0}^\infty v_i(y,t)\varphi ^{i-3}$
into (1)--(2), f\/ind the recursion relations for $u_i$ and $v_i$, and check
the compatibility conditions at the resonances. All the compatibility
conditions turn out to be satisf\/ied \textit{identically}. Thus, the analyzed
system has passed the Painlev\'e test well, and we should expect its
integrability.

Next we could see that the procedure of truncating singular expansions [6]
turns out to be compatible for the system (1)--(2). Unfortunately, the
explicit expressions for truncated expansions are too bulky, and the way
they lead to the Lax pair is too artif\/icial. We will show an easier way to
the same Lax pair. Here we only have to note that \textit{two} ``spectral''
parameters, $\alpha $ and $\beta $, appear in the truncated singular
expansions:
\begin{equation}
\alpha =\alpha (y):\forall\; \alpha ,\qquad \beta =\beta (x,y):\beta
_x=-\alpha _y.
\label{sakovich:3}
\end{equation}

Let us remind that the system (1)--(2) is generated by the KdV equation (1)
under the following perturbation (see [1] for details):
\begin{equation}
(x,t)\rightarrow \left( x,y,t\right) ,\qquad y=\varepsilon x,
\qquad u\rightarrow
u+\varepsilon v,\qquad o\left( \varepsilon ^2\right) .  \label{sakovich:4}
\end{equation}
The Lax pair for the KdV equation (1) is well known [7]:
\begin{equation}
\Phi _x+A\Phi =0,\qquad \Phi _t+B\Phi =0,\qquad
\Phi =\left(\begin{array}{c}\phi _1 \\[1mm] \phi _2 \ea \right),
\label{sakovich:5}
\end{equation}
\begin{equation}
A=\left(
\begin{array}{cc}
0 & u+\alpha  \\
-1 & 0
\end{array}
\right) ,\qquad B=\left(
\begin{array}{cc}
-u_x & u_{xx}+2u^2-2\alpha u-4\alpha ^2 \\
-2u+4\alpha  & u_x
\end{array}
\right) ,  \label{sakovich:6}
\end{equation}
$\alpha _x=\alpha _t=0.$ Let us see how the perturbation (4) acts on the Lax
pair (5):
\[
\Phi (x,t)\rightarrow \Phi \left( x,y,t\right) +\varepsilon \Psi \left(
x,y,t\right) ,\qquad \Psi =\left( \begin{array}{c} \psi _1 \\ \psi
_2\end{array} \right),
\]
\[
\Phi _x\rightarrow \Phi _x+\varepsilon \left( \Phi _y+\Psi _x\right) ,
\]
\[
A\rightarrow A+\varepsilon M,\qquad  \dim M=2\times 2,
\]
\begin{equation}
\Phi _x+A\Phi =0\rightarrow \left\{  \ba{l}
\Phi _x+A\Phi =0, \\[1mm]
\Psi _x+\Phi _y+A\Psi +M\Phi =0,
\ea  \right. \label{sakovich:7}
\end{equation}
\[
\Phi _t\rightarrow \Phi _t+\varepsilon \Psi _t,
\]
\[
B\rightarrow B+\varepsilon N,\qquad \dim N=2\times 2,
\]
\begin{equation}
\Phi _t+B\Phi =0\rightarrow \left\{
\ba{l} \Phi _t+B\Phi =0,\\[1mm]
\Psi _t+B\Psi +N\Phi =0.
\ea \right.  \label{sakovich:8}
\end{equation}
We can rewrite the linear systems (7) and (8) in the block form as follows:
\begin{equation}
\left(\begin{array}{c} \Phi \\ \Psi \end{array} \right)_x+\left(
\begin{array}{cc}
0 & 0 \\
1 & 0
\end{array}
\right) \left(\begin{array}{c} \Phi \\ \Psi \end{array}\right)_y +\left(
\begin{array}{cc}
A & 0 \\
M & A
\end{array}
\right) \left( \begin{array}{c} \Phi\\ \Psi \end{array}\right)
=0,  \label{sakovich:9}
\end{equation}
\begin{equation}
\left(\begin{array}{c} \Phi \\ \Psi \end{array} \right)_t+\left(
\begin{array}{cc}
B & 0 \\
N & B
\end{array}
\right) \left(\begin{array}{c} \Phi \\ \Psi \end{array} \right)
=0.  \label{sakovich:10}
\end{equation}
Now we can f\/ind the explicit form of $M$ and $N$, applying the perturbations
(4) and $\alpha \rightarrow \alpha +\varepsilon \beta $ to $A$ and $B$ (6):
\begin{equation}
M=\left(
\begin{array}{cc}
0 & v+\beta  \\
0 & 0
\end{array}
\right) ,  \label{sakovich:11}
\end{equation}
\begin{equation}
N=\left(
\begin{array}{cc}
-u_y-v_x & v_{xx}+2u_{xy}+4uv-2\alpha v-2\beta u-8\alpha \beta  \\[1mm]
-2v+4\beta  & u_y+v_x
\end{array}
\right) .  \label{sakovich:12}
\end{equation}
Question:
Should consider $\alpha $ and $\beta $ as constants? Answer: 
No, we have to take
the original conditions $\alpha _x=0$ and $\alpha _t=0$ and apply the
perturbation (4) to them, assuming that $\alpha (x,t)\rightarrow \alpha
(x,y,t)+\varepsilon \beta (x,y,t)$. In this way, we f\/ind that $\alpha
_t=0\rightarrow \alpha _t=\beta _t=0$ and $\alpha _x=0\rightarrow \left\{
\alpha _x=0,\beta _x=-\alpha _y\right\} $, i.e. we obtain exactly the
conditions (3) for the two spectral  parameters $\alpha $ and $\beta $.
Moreover, we can check directly that the system (1)--(2) follows (as the
compatibility condition) from the linear systems (9) and (10) if, and only
if, $\alpha $ and $\beta $ satisfy the conditions (3). Thus, the system
(1)--(2) possesses the two-parameter 4$\times $4 Lax pair
\begin{equation}
\Omega _x+S\Omega _y+P\Omega =0,\qquad \Omega _t+Q\Omega =0  \label{sakovich:13}
\end{equation}
with the notations evident from (9), (10), (6), (11), (12) and (3). It
remains a question whether some other remarkable objects of the system
(1)--(2), such as solitons, conservation laws, as well as Miura and B\"acklund
transformations, can also be derived \textit{simply by applying the
perturbation}~(4) to the corresponding objects of the KdV equation
(1).

The system (1)--(2) admits two interesting (1+1)-dimensional reductions. If
we make $\left( x,y,t\right) \rightarrow (z,t)$ and choose $z=x$, we get
from (1)--(2) the well-known integrable perturbed KdV equation [1, 8]
\begin{equation}
u_t=u_{zzz}+6uu_z,\qquad v_t=v_{zzz}+6(uv)_z.  \label{sakovich:14}
\end{equation}
Choosing $z=x+y$ and introducing $w=u+v$ for simplicity, we get the new
system
\begin{equation}
u_t=u_{zzz}+6uu_z,\qquad  w_t=w_{zzz}+6(uw)_z+3u_{zzz}.  \label{sakovich:15}
\end{equation}
The Lax pair of (15) follows directly from (13) by the reduction. Moreover,
since the systems (14) and (15) came from a system possessing the Painlev\'e
property by reductions to its  non-characteristic hypersurfaces, they
possess the Painlev\'e property automatically. This is essential in relation
with one recent Painlev\'e classif\/ication of coupled KdV equations [9]:
neither (14) nor (15) appeared there as systems that had passed the
Painlev\'e test for integrability. Consequently, the classif\/ication [9]
should be revised. This work is in progress.

\label{sakovich-lp}

\end{document}